\documentclass[twocolumn]{article}
\pdfoutput=1
\newif\ifpdf\ifx\pdfoutput\undefined\pdffalse\else\pdfoutput=1\pdftrue\fi

\usepackage{amsmath,amssymb,graphicx}
\begin{document}

\title{ Adaptive  Multi-Dimensional Particle In Cell}
\author{Giovanni Lapenta \\ Centre for Plasma Astrophysics, Departement Wiskunde, \\ Katholieke Universiteit Leuven\\ Celestijnenlaan 200B, BE-3001 Heverlee, Belgium  \\(giovanni.lapenta@wis.kuleuven.be)}

   \maketitle      
\begin{abstract}Kinetic Particle In Cell (PIC) methods can extend greatly
their range of applicability if implicit time differencing and
spatial adaption are used to address the wide range of time and
length scales typical of plasmas. For implicit differencing, we
refer the reader to our recent summary of the implicit moment PIC
method implemented in our CELESTE3D code [G. Lapenta, Phys. Plasmas,
13, 055904 (2006)]. Instead, the present document deals with the issue of PIC
spatial adaptation. Adapting a kinetic PIC code requires two tasks:
adapting the grid description of the fields and moments and adapting
the particle description of the distribution function. Below we
address both issues. First, we describe how grid adaptation can be
guided by appropriate measures of the local accuracy of the
solution. Based on such information, grid adaptation can be obtained
by moving grid points from regions of lesser interest to regions of
higher interest or by adding and removing points. We discuss both
strategies. Second, we describe how to adapt the local number of
particles to reach the required statistical variance in the
description of the particle population. Finally two typical
applications of adaptive PIC are shown: collisionless shocks and
charging of small bodies immersed in a plasma. 
\end{abstract}

\section{Introduction}

Methods to adapt particle-in-cell (PIC) kinetic plasma
calculations are very valuable in the study of multiple-length
scale problems. Typically, multiple-length scale problems present
small regions of stronger gradients embedded in large systems. In
such conditions, computational efficiency is achieved best by
focusing the attention in the regions of interest.

In PIC methods it is not sufficient to use adaptive grids with finer
spacing in the regions of interest, it is also necessary to rezone
the number of particles. By particle rezoning we define the
operation of increasing the number of particles in regions where
higher accuracy is required, and of reducing the number of particles
where lower accuracy can be tolerated. Finer grid spacing leads to a
better description of the electromagnetic fields, but particle
rezoning is needed to gain a better description of the plasma
dynamics and a reduction of noise~\cite{lapenta94}.

Particle rezoning can also be beneficial to keep the load of work
uniform on a per cell basis, a feature of crucial interest in a
correct load balancing in parallel implementations.

In the present work, we review our work in the field, without
attempting to present a complete coverage of the literature. The
paper is organizes as follows.

Section 2 reports general comments regarding the task of PIC
adaption, discussing in particular the link of particle and grid
adaption and the link of both with time differencing. Section 3
deals with grid adaptation, while Section 4 deals with particle
adaptation. Section 3 is organized around the two main task of grid
adaption: where to adapt and how to adapt. We answer the question of
where to adapt by proposing a posteriori measures of the local
accuracy. We answer the question of how to adapt by examining two
different possibilities: grid motion and grid refinement.

Finally, Section 5 presents to types of simulations where PIC
adaption is tested: shocks and dust charging.

\section{PIC Code Adaptation}

Multiple scale plasma physics problems present two challenging
features. First, in any given region of the system processes develop
at widely different time scales. Electrons and ions respond with
scales made extremely different by their different masses, and a
host of different intabilities can develop each with its own time
and length scales. Second, different regions of the system can have
widely separated spatial and temporal scales. For examples, regions
of localized strong gradients can arise locally, as is the case of
current sheets in space systems.

The normal textbook approach to PIC is unsuitable to the conditions
described above. The standard PIC is based on explicit time
differencing and is subject to strict stability constraints. The
time step needs to resolve both light-wave propagation and Langmuir
wave propagation:
\begin{equation}
c \Delta t<\Delta x \label{finitec}
\end{equation}
\begin{equation}
\omega_{pe}\Delta t<2 \label{finitep}
\end{equation}
 regardless to our relevance to the scale of
interest. The grid spacing needs to resolve the electron Debye
length:
\begin{equation}
\Delta x< \varsigma \lambda_{De} \label{finitegrid}
\end{equation}
to avoid the so-called finite grid instability~\cite{birdsall}. For
this reason explicit methods need to resolve the finest scales
everywhere.

When implicit methods are considered~\cite{brackbill-forslund}, the
stability constraints (\ref{finitec}, \ref{finitep}, \ref{finitegrid}) are removed and the local
spacing can be chosen according to the required accuracy rather than
the need to avoid  instability. A practical condition that ensure good energy conservation in an implicit PIC method requires that the average electron population does not travel more than one cell per time step:
\begin{equation}
v_{th,e}\Delta t<\Delta x
\end{equation}
This condition can be satisfied if both grid spacing and time step (but not the one only) are chosen large, to step over the small and fast scales. The scales not resolved accurately are not eliminated but rather deformed and the energy present in them is damped like in the physical Landau damping but at an accelerated rate \cite{brackbill-forslund}.

We refer the reader
to our recent review of the implicit PIC algorithm used in the
CELESTE code for the details of how the implicit moment method is
derived and used~\cite{lapenta05,brackbill-forslund}. In the present paper we describe,
instead, how implicit PIC methods can be adapted in space.

To adapt a PIC code to a local scale length, we need to address two
issues, how to change the local grid resolution and how to change
the local statistical description of the particle distribution
functions. The two issues are described in the two sections below. The assumption is made that the 
host PIC method be implicit, so that large cells do not lead to the finite grid instability described above. Nevertheless, some forms of adaptation can also be of relevance to explicit methods and some of the methods described below can also be used in explicit codes (for example the application in section 5.2 is explicit).

\section{Grid Adaptation}

Grid adaptation can be achieved by grid refinement (i.e. adding more
grid points) in some selected areas or by grid motion (i.e. moving
grid points to regions of interest from regions of lesser interest).
In the first case, the adaptive mesh refinement (AMR)
method~\cite{berger} is obtained. In the second case, the moving
mesh adaptation (MMA) method~\cite{jerrygrid1}  is obtained. A
specific class of MMA algorithms widely used are ALE
methods~\cite{hirt}. In all cases we need guidance. We need to know
what interesting mean. Often, interest is defined based on the
knowledge of the solution.

In many plasma physics problems the regions of interest are readily
identified. For example, in space weather simulations localized
regions of strong currents are site of topological changes and
require a high resolution while regions of smooth flow can be
described by coarse meshes.

 However, in other instances it is not obvious what regions require
 increased accuracy. In those cases, we need error detectors to tell us where the error
is larger. Here we describe a specific error detector previously
applied successfully in plasma physics problems: the operator
recovery error origin (OREO) detector~\cite{lapentaijnme}.

 For AMR codes, the OREO detector
 provides accurate and automatic determination of where
the discretization error is being generated. This knowledge is
directly used by the AMR method to refine or to coarsen.

For MMA codes, the knowledge of the error needs to be supplemented
by a method to move the grid. Given the error what new grid should
we use? To answer this additional problem typical of the MMA method
we also present a new technique~cite{lapenta} based on the
Brackbill-Saltzman approach~\cite{jerrygrid1}.

\subsection{Automatic Guidance on Resolution Requirements}

In a previous paper \cite{lapentaijnme}, we have proposed a new
error origin detector based on the extension of the gradient
recovery error estimator~\cite{ainsworth}. We have named the
approach {\it operator recovery error origin (OREO) detector }
since it extends to any operator the method used for the gradient
operator by the gradient recovery error estimator. Below, we
summarize briefly the procedure involved in its definition and
implementation.

For the sake of definiteness, we shall assume a general
N-dimensional grid (where one of the dimensions could be time)
where a vector field ${\bf v}_n$ is node centered. For notation,
we label the cells with $c$ and the nodes with $n$, using further
the notation $n(c)$ to indicate the nodes neighboring cell $c$ and
$c(n)$ to indicate the cells neighboring
 node $n$.

 We consider a general
multi-dimensional non-linear partial differential operator:
\begin{equation}
{\mathbb O}(q)\label{operator}
\end{equation}
Equation (\ref{operator}) summarizes the most general operator
acting on a function $q({\bf x})$ defined on the multidimensional
space ${\bf x}$.

Equation (\ref{operator}) is discretized  on a grid with $N$ nodes
${\bf x}_n$:
 \begin{equation}
O_n (q_1,\ldots,q_N)\label{mfesampledisc}
\end{equation}

From the discretized field $q_n$ and from the discretized operator
$X_{n}$  applied to $q_n$ defined only on the grid nodes, it is
possible to reconstruct two functions defined everywhere in the
continuum space ${\bf x}$:
\begin{equation}
\begin{array}{l}
\widetilde{q}({\bf x})=\sum_n q_n S({\bf x}-{\bf x}_n) \\
\widetilde{O}({\bf x})=\sum O_{n} S({\bf x}-{\bf x}_n)\\
\end{array}
\end{equation}
where $S({\bf x}-{\bf x}_n)$ is the b-spline basis function of order
$\ell$ for interpolation.

The local truncation error is  defined as the difference between
the linear interpolation of the discretized operator applied to
the discretized field  $\widetilde{X_q}({\bf x})$ and the exact
differential operator applied to the linear interpolation of the
discretized field  $\widetilde{q}({\bf x})$:
\begin{equation}
e= \widetilde{O}({\bf x})- {\mathbb O} \widetilde{q}({\bf
x})\label{trunc}
\end{equation}

The average local truncation error on any given cell $c$ is defined
as a norm of the error $e$. The $L_2$ norm is often used:
\begin{equation}
{e}_c =\left(\frac{1}{V_c}\int_{V_c} e^2 dV
\right)^{1/2}\label{lsmfe}
\end{equation}
where ${e}_c$ is the average local truncation error over cell $c$
and $V_c$ is the cell volume.

\subsection{Moving Mesh Grid Adaptation}

We have recently proposed a new approach~\cite{lapentagrid} to
variational grid adaptation~\cite{jerrygrid1} based on the
minimization of the local truncation error defined above. The method
can be constructed starting from the following equidistribution
theorem proven in Ref.~\cite{lapentagrid}

 {\sc theorem}: {\it In a
optimal grid, defined as a grid that minimizes the local
truncation error according to the minimzation principle
\begin{equation} \int_{\mathbb V} \mid e \mid d^Nx\,,
\label{minxnd}
\end{equation}
the product of the local truncation error in any cell $i$ by the
cell volume $V_i$ (given by the Jacobian $J=\sqrt{g}$) is
constant:
\begin{equation}
 e_i V_i ={\rm const}\label{const1d4}
\end{equation}
}

The equidistribution theorem is applied solving the following
Euler-Lagrange equations:
\begin{equation}
g^{ij} \frac{\partial }{\partial \xi^i} \left(|e| \frac{\partial
x^i}{\partial \xi^j}\right)=0 \label{elqui}
\end{equation}
This approach creates a grid where $|e_{i}| V_i$ is constant. Note
that the equations above are identical to the equations used by the
Brackbill-Saltzman variable diffusion method~\cite{jerrygrid1}. The
primary innovation is that the monitor function is now directly
linked with the local truncation error instead of being left
undefined. In the typical implementations of the Brackbill-Saltzman
method, the monitor function is defined heuristically by the user.
The use of the OREO detector proposed here results in a more
accurate scheme~\cite{lapentagrid}.

We have applied the grid rezoning described above to our MMA
magnetohydrodynamics (MHD) code GRAALE~\cite{giannifinn} based on
the ALE discretization~\cite{hirt}. Here we limit the discussion to
the classic spherical 1D implosion test proposed by Noh~\cite{noh}.
An unmagnetized gas with $\gamma=5/3$ initially has $\rho=1$,
$e=10^{-4}$ and uniform velocity $u=-1$ (except in the center where
$u(r=0)=0$). The problem represents a serious challenge for
Lagrangian calculations and the solution is known to suffer from
serious wall heating due to the use of artificial viscosity to
capture shocks. Note that we are not using artificial heat
conduction~\cite{noh} (a tool to mitigate the wall heating problem)
precisely to highlight the trouble of Lagrangian calculations for
the present case .

\begin{figure}
\includegraphics[width=80mm,angle=0]{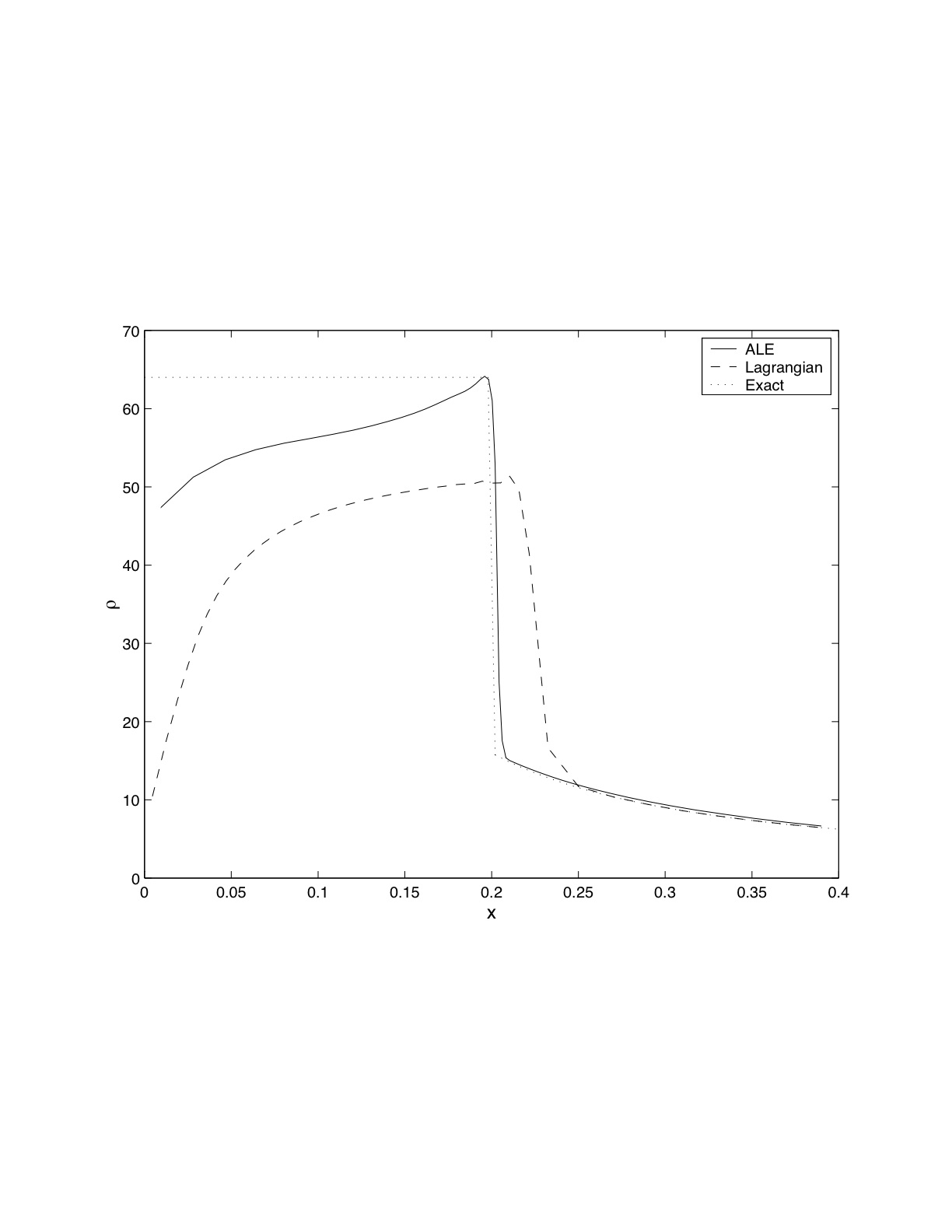}
\caption{ Noh's spherical benchmark: comparison of the density at
the end ($t=0.6$), for a  Lagrangian (dashed) and MMA (solid line)
calculation. The exact solution is also shown (dotted line).}
\label{noh}
\end{figure}

 The results of an MMA calculation using the adaptive grid
 is compared with a reference standard Lagrangian
calculation. Figure \ref{noh} shows the density at the end of the
Lagrangian and MMA calculation. The use of adaptive grid results in
a much improved solution. The reason for the improvement is
explained by the sharper resolution  of the shock achieved by the
adaptation. As noted in the original paper by Noh~\cite{noh}, a
sharper resolution of the shock also implies a reduction of wall
heating, as observed in Fig.~\ref{noh} for the MMA case.

The use of  grid adaptation based on the OREO detector results in an
automatic method to increase the accuracy of the MMA method.

\subsection{Adaptive Mesh Refinement (AMR) }

\begin{figure}
\centering
\includegraphics[width=80mm,angle=0]{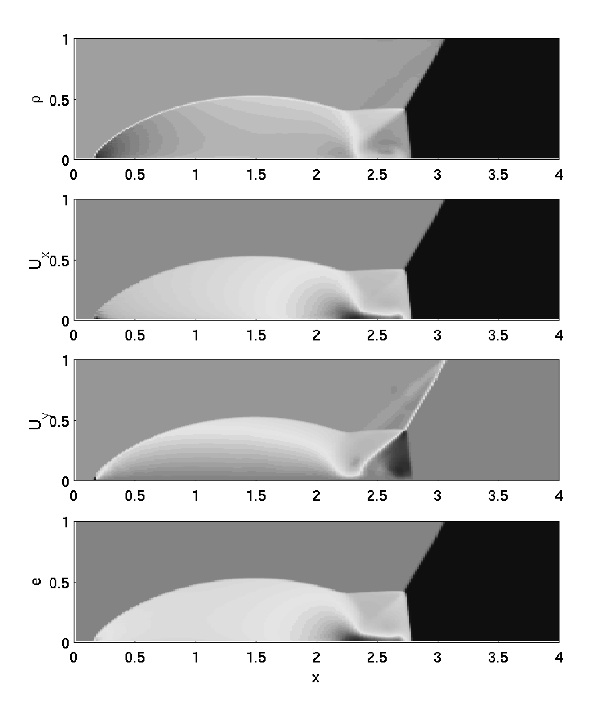}
 \caption{2D Gas Dynamics (Eulerian form)- Colella's benchmark on a
 240x120grid. Density, velocity and internal energy at the
end  of a Eulerian calculation ($t=0.2$). Results obtained using
CLAWPACK. } \label{rescolella}
\end{figure}

To investigate the performance of the  OREO detector in 2D, we have
applied it to results obtained with CLAWPACK~\cite{leveque}.
CLAWPACK is a publicly available software based on an AMR
solution~\cite{berger} of the conservation laws. We have applied the
code to the solution of the gas dynamics equations for the Colella's
wedge problem~\cite{colella}. A planar $M=10$ shock is incident on
an oblique surface; the angle between the shock direction and the
surface is $\pi/6$.
 The actual computed results at time $t=0.2$
for a 240x120 grid are shown in Fig.~\ref{rescolella} where all
the expected features~\cite{colella} can be recognized.

\begin{figure}[htb]
\centering
 \includegraphics[width=80mm,angle=0]{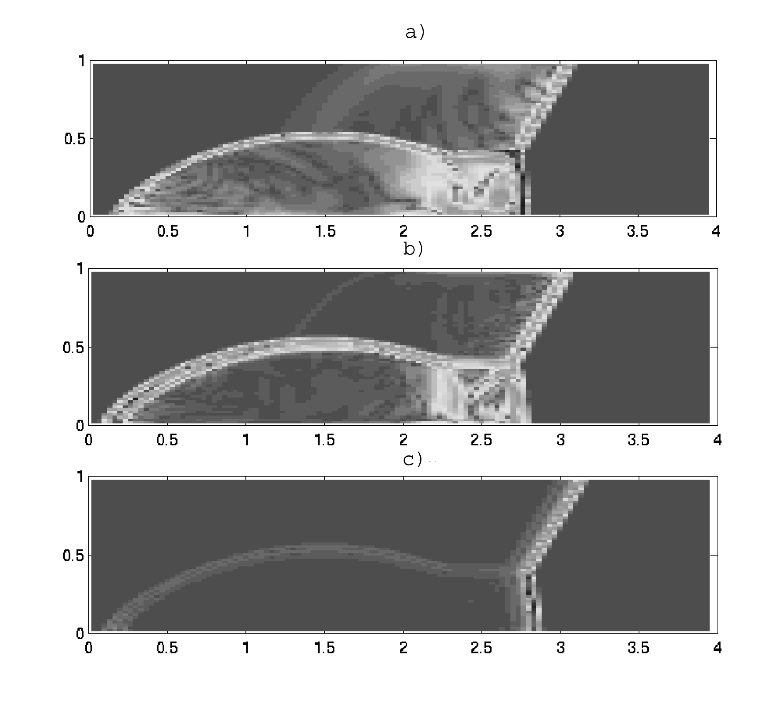} \caption{2D Gas
Dynamics - Colella's benchmark on a
 120x60 grid at $t=0.2$. Comparison  of
the global truncation error (a) with the OREO detector (b), and
the warp indicator (c). } \label{detcolella}
\end{figure}

The OREO detector is computed based on the results obtained from
CLAWPACK using Algorithm 3. The detector is shown in
Fig.~\ref{detcolella} for a simulation with a grid 120x60. For
comparison we also provide an estimate of the actual error,
computed by difference between the solution on a 120x60 grid and
the more accurate solution on a 240x120 grid. Clearly the OREO
detector is successful in detecting all origins of errors. The
shocks are all captured; the slip surface rolling up under the
shock is evident. All features are detected.

For reference, Fig.~\ref{detcolella}-c shows also a similar
analysis conducted on another possible candidate for error
detection often used in the literature. The detector, which we
name {\it warp indicator} for convenience, measure the local error
as the variance among the different values obtained at a node when
extrapolating the internal energy from the four directions
(backward and forward along $x$ and backward and forward along
$y$). The analysis in Fig.~\ref{detcolella}-c shows that the two
rightmost planar shocks are captured well, while the top and bow
shocks are barely visible. All the structure inside the rolling up
region within the outer shocks is lost: no slip surface is
measured and the internal shock is also lost. In practice the warp
indicator is often supplemented by other ad hoc detectors to pick
up all shocks, but still the rolling up region and the slip
surfaces are often left undetected.

The OREO detector does not miss any feature and can be used
reliably alone without any other ad hoc detector.

\section{Particle Adaptation}
Particle rezoning is needed to increase the number of particles in
regions where high accuracy is required, and to reduce the number
of particles where lower accuracy can be tolerated. The primary
effect of increasing the number of particles is to reduce the
variance of the statistical description of the  distribution
function. In a PIC simulation this increases the accuracy defined
as typical in MonteCarlo methods, i.e. as the variance of the
simulation.

Particle rezoning must be in effect throughout the calculation to
constantly keep the local required accuracy. In multiple-length
scale problems, the region of interest can move, and particle
rezoning must follow the motion to keep the focus where it is
needed. The approach followed here is to use adaptive grids to
follow the evolution of the system~\cite{jerrygrid1, lapentagrid}
and particle rezoning to keep the number of particles per cell
constant. This approach leads to finer grid spacing in the region of
interest and, automatically, to a higher density of computational
particles in that region.

The problem of particle rezoning can be formulated \cite{lapentaadapt} as the
replacement of a set of $N$ particles with position ${\bf x}_p$,
velocity ${\bf v}_p$, charge $q_p$, and mass $m_p$, with a
different set of $N^{\prime}$ particles with
position ${\bf x}_{p^{\prime}}$, velocity ${\bf v}_{p^{\prime}}$, charge $%
q_{p^{\prime}}$, and mass $m_{p^{\prime}}$. The criterion for
replacement is the {\it equivalence\/} between the two sets,
defined as the requirement that the two sets must represent the
same physical system, with a different accuracy. This generic
definition of equivalence between two sets is given practical
bearing by specifying two rules for equivalence.

Two sets of particles are considered equivalent if \cite{lapentaadapt}:

\begin{enumerate}
\item  the two sets are indistinguishable on the basis of their
contributions to the  grid moments;

\item  the two sets of particles sample the same velocity
distribution function.
\end{enumerate}

The first criterion concerns the moments of the particle
distribution used
to solve the field equations. The moments are defined at the grid points $%
{\bf x}_g$ as
\begin{equation}
M_g = \sum\limits_p S \left({\bf x}_g - {\bf x}_p\right)q_p {\bf F} ({\bf v}%
_p)\;,
\end{equation}
where $S$ is the assignment function~\cite{hockney, bspline}. In
general, when nonuniform grids are used, {\bf x}\ is the natural
coordinate, i.e. the system of coordinates where the spacing between
consecutive points is uniform and unitary in all
directions~\cite{jerrygrid1}. The function {\bf F} of the particle
velocity characterizes the moment. In explicit electrostatic codes,
only the charge density is required:
\begin{equation}
\rho_g = \sum\limits_p S_g ({\bf x}_p) q_p
\end{equation}
derived from~(1) using ${\bf F} ({\bf v}_p) = 1$ and using a short
notation for $S_g ({\bf x}_p) = S ({\bf x}_g - {\bf x}_p)$.
Electromagnetic and implicit codes~\cite{vu} require higher order
moments like the current density
\begin{equation}
{\bf J}_g = \sum\limits_p S_g ({\bf x}_p)q_p {\bf v}_p
\end{equation}
and the pressure tensor
\begin{equation}
\Pi_g = \sum\limits_p S_g ({\bf x}_p)q_p {\bf v}_p {\bf v}_p\;.
\end{equation}

The first criterion requires the two sets of particles to give the
same moments relevant to the field equations. Note that if this
criterion is satisfied exactly total energy and momentum are also
automatically conserved. The second criterion is more difficult to
apply in a quantitative fashion. In previous work~\cite
{lapenta94,lapentaadapt}, it has been proposed to use the $\chi^2$ test or the
Kolmogorov and Smirnov test to verify that the particle
distribution is preserved. In practice, this is not easily
achieved.

In fluid PIC codes, the first criterion is the only one to be
applied, and general schemes for particle rezoning can be
derived~\cite{lapenta95}. In kinetic PIC codes, the computational
particles sample the real plasma velocity distribution, and the
second criterion must also be imposed. In the kinetic case the
choices are more limited. For this reason, a simpler approach is
followed~\cite{lapenta94,lapentaadapt}. To increase the number of particles per
cell, a given particle is split in two or more new particles
displaced in space but all sharing the same speed. The weights and
displacements can be chosen to conserve exactly the grid moments,
and the velocity distribution is not altered because all the
particles have the same velocity.

Another approach can be considered. A particle can be split in the
velocity space. The daughter particles have the same position but
different velocity. The advantage of this method is that the charge
density is not affected. However, the higher order moments (current
density and energy) cannot be all preserved. Furthermore, the
velocity distribution is altered.

To decrease the number of particles, the splitting operation can
be inverted to coalesce two particles into one. The difficulty is
that, in general, it is impossible to find two particles with the
same velocity. For this reason, particles with different velocity
have to be coalesced. To minimize the perturbation of the velocity
distribution, the particles to be coalesced must be chosen with
similar velocity. An alternative approach is to coalesce three
particles into two, which allows one to conserve both energy and
momentum~\cite{vahedi}.

In the following sections, we will provide the two most successful general techniques to
adapt the number of particles in a cell. We refer the reader to a previous technical description of the various alternatives and their merits~\cite{lapentaadapt}

\subsection{Summary of the Algorithms for Particle Rezoning}

\nobreak In the previous sections, we derived all the required
blocks to build algorithms to change the number of particles in any
given cell. Here, we provide a precise algorithmic description of
the methods to increase the number of particles per cell and to
decrease the number of particles per cell.

\noindent {\bf Splitting Algorithm }:

\nobreak {\it Given a cell $g$ with $N_p$ particles in a 1D, 2D,
or 3D system, any chosen particle (labeled $o$) with charge $q_o$
(and mass obtained from the charge-to-mass ratio for the species),
position ${\bf x}_o$ (in natural coordinates) and velocity ${\bf
v}_o$ can be replaced by $N^{\prime}$
particles, labeled $p^{\prime}= \{1, 2\ldots N^{\prime}\}$. In 1D, $%
N^{\prime}= 2$ and the new properties are $q_{p^{\prime}} = q_o/2$, $%
x_{p^{\prime}}=x_o \pm 1/N_p$ (where the cell size is unitary), ${\bf v}%
_{p^{\prime}} = {\bf v}_o$. In 2D, $N^{\prime}= 4$ and the new
properties
are $q_{p^{\prime}}=q_o/4$; $x_{1,2}= x_o \pm 1/N_p$, $x_{3,4}=x_o$, $%
y_{1,2} = y_o$, $y_{3,4}= y_o \pm 1/N_p$; ${\bf v}_{p^{\prime}} = {\bf v}_o$%
. In 3D, $N^{\prime}= 6$ and the new properties are $q_{p^{\prime}}=q_o/6$, $%
x_{1,2}=x_o \pm 1/N_p$, $x_{3,\ldots 6} = x_o$; $y_{1,2,5,6} = y_o$, $%
y_{3,4} = y_o \pm 1/N_p$; $z_{1,\ldots 4} = z_o$; $z_{5,6} = z_o
\pm 1/N_p$. }

Note that the choice of the particle $p=o$ in the set of $N_p$
particles in the cell $g$ is free. In the result sections, we
choose the particle with the largest energy: $m_p{\bf v}^2_p$.
Algorithm~S1 preserves exactly the velocity distribution function
and grid moments. However, for quadratic assignment functions the
grid moments are only approximately preserved (see Section~III).

\noindent {\bf Coalescence Algorithm}:

\nobreak
{\it Given a cell $g$ with $N_p$ particles in 1D, 2D, or 3D systems, choose $%
N = 2$ particles $p = \{1,2\}$ close to each other in the phase
space. Their properties are $q_p$, ${\bf x}_p$, and ${\bf v}_p$.
The two chosen particles can
be replaced by one particle (labeled $A$) with $q_A = q_1 + q_2$, ${\bf x%
}_A = (q_1{\bf x}_1 + q_2{\bf x}_2)/q_A$, ${\bf v}_A = (q_1{\bf
v}_1 + q_2 {\bf v}_2)/q_A$. }

Algorithm~C1 preserves the overall charge and momentum and the
charge density $\rho _g$ but perturbs the velocity distribution.
Note that one can choose ${\bf v}_A$ to preserve the energy, but
it is not possible to preserve energy and momentum together. The
crucial point of algorithm~C1 is to choose two particles close in
velocity and space. A pair search of the two particles closest in
velocity is usually too expensive. For this reason, we perform a
diatomic search that sorts the particles into two bins and selects
the largest bin. The binning is repeated in sequence for each
spatial direction and component of the velocity. The binning is
continued until the number of particles in the largest bin is
small enough to use a pair search.

\section{Examples of Adaptive PIC Simulations}

To illustrate the possible applications of adaptive PIC method,
below we report two classic cases where uniform PIC calculations
show their limitations: collisionless shocks and small scales
objects (dust particles) immersed in plasmas.

\subsection{Collisionless Shocks}

\nobreak Simulations of collisionless shocks provide a sensitive
test of the accuracy of particle rezoning
methods~\cite{lapenta94}. In the slow shock calculations
considered here, a magnetized plasma is flowing toward a piston
that reflects the particles. A switch off slow shock is
considered, and the component of the magnetic field perpendicular
to the normal of the piston is set to zero.

We consider here the same conditions reported in
Ref.~\cite{lapenta94}. The initial configuration is chosen according
to the Rankine-Hugoniot conditions. The initial ratios of the
electron and ion pressures to the upstream magnetic field are
$\beta_e = \beta_i = 0.01$. The ratio of ion to electron mass is
$m_i/m_e = 25$; the ratio of the upstream ion cyclotron and ion
plasma frequencies is $\omega_{ci}/\omega_{pi} = 0.01$, and the
shock normal angle, with respect to the magnetic field, is $\psi =
75^\circ$. The size of the simulation region is $L =
200\,c/\omega_{pi}$, and the shock is followed until $\omega_{ci} t
= 50$. Particles are injected at the right boundary to simulate a
flowing plasma.

The simulations are performed using CELESTE1D~\cite{vu}, a 1D
implicit PIC code, suitably modified by the author to include
particle control.

As a reference, we conduct a reference collisionless shock
calculation with a uniform grid and without particle rezoning. The
grid has 1000 cells giving a uniform spacing with $\Delta x =
0.2\,c/\omega_{pi}$; 128 electrons and 128 ions per cell are used.
Figure~4 shows the stack plot of $B_z$ as a function of the
position at 50 equally spaced time intervals between $t = 0$ and $%
\omega_{ci} t = 50$.

\begin{figure}[htb]
\centering
 \includegraphics[width=80mm,angle=0]{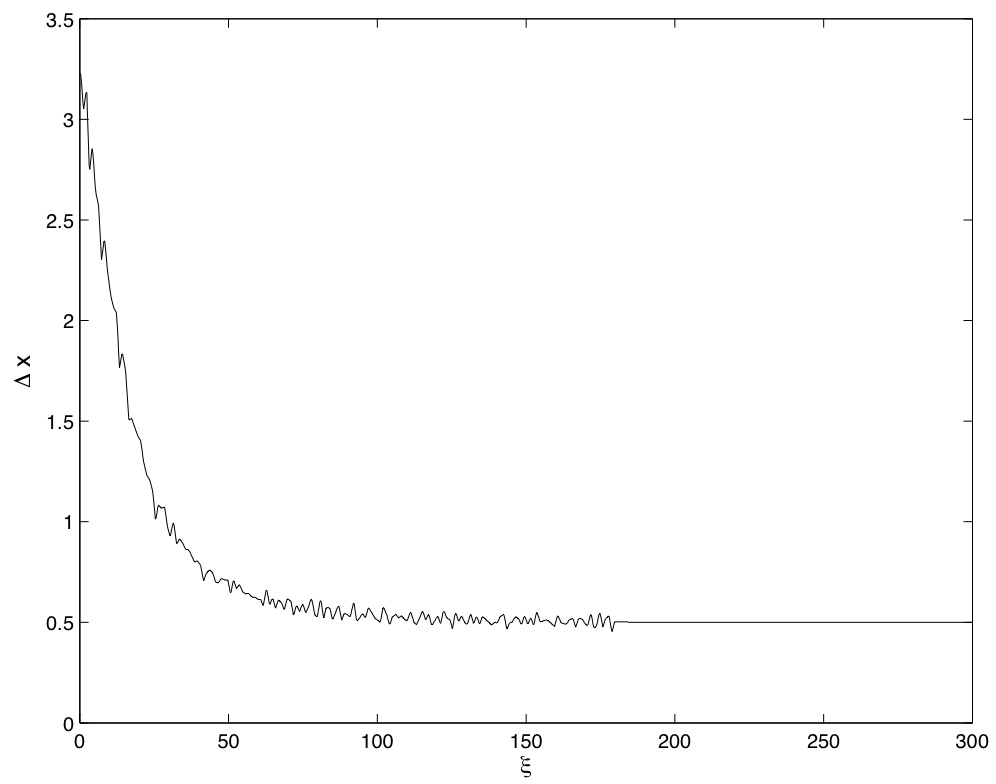} \caption{Grid used in the adaptive shock
 calculation. The grid size in each cell is plotted versus the cell
 center.
 The region of smaller grid spacing moves to the left to follow the shock.} \label{grid}
\end{figure}

The reference results are compared with a calculation where particle
rezoning is performed using the algorithms described above for
splitting and for coalescences. The computation uses an adaptive
grid with finer spacing in the shock region ($\Delta x \approx
0.5\,c/\omega_{pi}$) and coarser outside. The region of fine spacing
expands in time to follow the motion of the shock. The grid spacing
in the region of the shock is kept fixed; and, consequently, the
grid spacing in the coarser region grows to keep the number of grid
points constant and equal to 300. Figure~\ref{grid} shows the grid
spacing at the end of the simulation, $\omega_{ci t}=50$. Note that
the area of the shock is well resolved, while the upstream region
has large cells. We use the grid jiggling technique of randomly
displacing the grid spacing in the large cells to improve the energy
conservation of the simulation~\cite{jiggling}. This technique
results in a random noise added to the grid spacing in the large
cells. To avoid any noise in the shock region, the jiggling
technique is not used there.

The particles are loaded with a uniform number per cell (the same as
before), leading to higher accuracy where the grid is finer.
Particle rezoning is required to keep the uniformity of the number
of particles per cell as the grid is adapted.

\begin{figure}[htb]
\centering
 \includegraphics[width=80mm,angle=0]{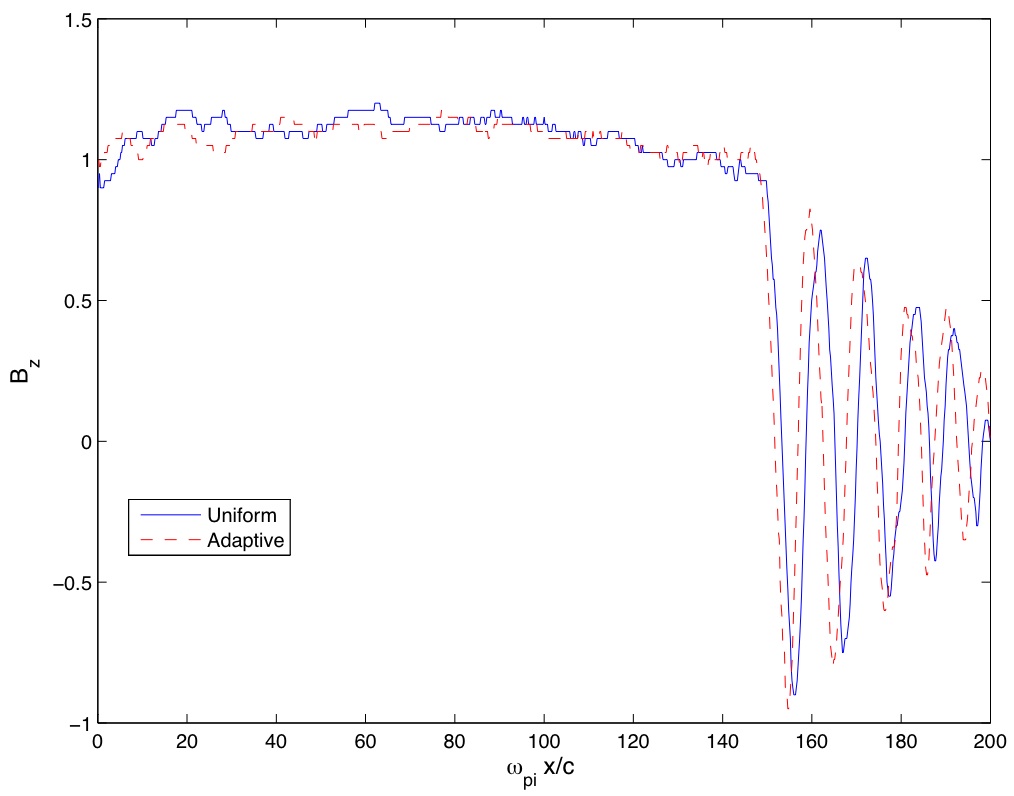} \caption{ Spatial profile of the $z$ component of the magnetic
 field $B_z$ at the end of the simulation (normalized to its upstream value) at time $\omega_{ci t}=50$. Two different
 runs are shown. The first uses a uniform 1000 cells grid (solid line) and the second an adaptive 300 cells grid (dashed line).} \label{shock}
\end{figure}

Figure~\ref{shock} shows the  profile of $B_z$ at the end
($\omega_{ci t}=50$) of the two
 calculation described above. Clearly, the evolution of the
system is calculated correctly. In particular,
the shock has traveled backward along the axis for a length of $%
50\,c/\omega_{pi}$ as in the reference case (Fig.~4) and as
required by the Rankine-Hugoniot conditions. The results have been validated against previously published results~\cite{vushock}.

\subsection{Charging Of Dust Particles}

As a second test, we consider small objects (e.g. dust particles)
immersed in a plasma. This condition is common in industrial
applications of plasma physics and in space and astrophysical
occurrences of dusty plasmas. Dust particles immersed in plasmas
tend to acquire a negative charge. The ions and electrons of the
plasma reach the surface and stick to it. If no secondary emission
or photoemission is present, the equilibrium charge on the dust
particle must be negative to repel the more mobile electrons and
attract the ions to achieve a balance of electron and ion currents.
This problem is of interest in laboratory and in space plasmas~\cite
{popdust99,prldust95}. We consider here the case where a plasma with
an ion to electron temperature ratio $T_e/T_i = 20$ and ion to
electron mass ratio $m_i/m_e = 1836$ is drifting relative to a
spherical dust particle of radius $a/\lambda_{De} = 0.4$, where
$\lambda_{De}$ is the electron  Debye length. The relative velocity
$w$ is expressed by the Mach number $M = w m^{1/2}_i/(kT_e)^{1/2} =
10$. The system is simulated using a cylindrical coordinate system
with the vertical axis along the direction of the plasma flow and
centered in the center of the spherical dust particle. In this
configuration, the azimuthal coordinate is invariant, and the
problem is 2D axisymmetric.

The interaction of the dust particle with the plasma is described
with the immersed boundary method. The application of the immersed
boundary method in
PIC codes is described in Ref.~\cite{sulsky} for fluid problems and in Ref.~%
\cite{lapentaieee} for plasmas. In the present work, we will use the
immersed boundary explicit PIC code DEMOCRITUS developed by the
author for dusty plasma simulations~\cite{popdust99}. A brief
description of the method is given below, more details can be found
in Ref.~\cite{popdust99}.

The dust particle is represented by motionless computational
particles (object particles) with properties suitable to describe
the macroscopic properties of the dust. Dust plasma interface
conditions are treated with the immersed boundary method in two
steps.

First, we assign to the object particles a susceptibility $\chi_p$
that can be interpolated to the vertices of the grid ${\bf x}_v$
to obtain a grid susceptibility:
\begin{equation}
\chi_v = \sum_p S_{vp} \chi_p\;,
\end{equation}
where $S_{vp}$ are the linear assignment weights. The grid
susceptibility is used to alter the Poisson's equation:
\begin{equation}
{\cal D}_{cv} (1 + \chi_v) {\cal G}_{vc^{\prime}}
\phi_{c^{\prime}} = \rho_c\;,
\end{equation}
where the potential $\phi$ and the charge density $\rho$ are
defined on the
cell centers ${\bf x}_c$ and repeated indexes are summed. The operators $%
{\cal D}_{cv}$ and ${\cal G}_{vc}$ are a difference approximation
of the divergence and gradient, respectively. As discussed in
detail elsewhere~\cite {sulsky,lapentaieee}, Eq.~(23) is solved
everywhere, including in the interior of the dust particle. The
term $(1+\chi_v)$ gives an approximation to the correct interface
conditions for the electric field. In the present case, $\chi_v$
is the susceptibility of dielectric dust.

Second, the object particles exert a friction on the plasma
particles, via a slowing property $\mu_p$ that is interpolated to
the grid, as in Eq.~(22), to produce a grid quantity $\mu_v$ used
to introduce a damping term to the equation of motion of the
plasma particles:
\begin{equation}
{\frac{d{\bf v}_p }{dt}} = \sum_v {\bf E}_v S_{vp} - {\bf v}_p
\sum_v S_{vp} \mu_v\;.
\end{equation}
The second term in Eq.~(24) can be as big as desired to stop the
plasma particles on the surface of the dust. The damping term is
zero everywhere outside the region occupied by the dust.
Equation~(23) and Eq.~(24) allow one to treat the field and
particle boundary conditions on the surface of the dust.

\begin{figure}[htb]
\centering
 \includegraphics[width=80mm,angle=0]{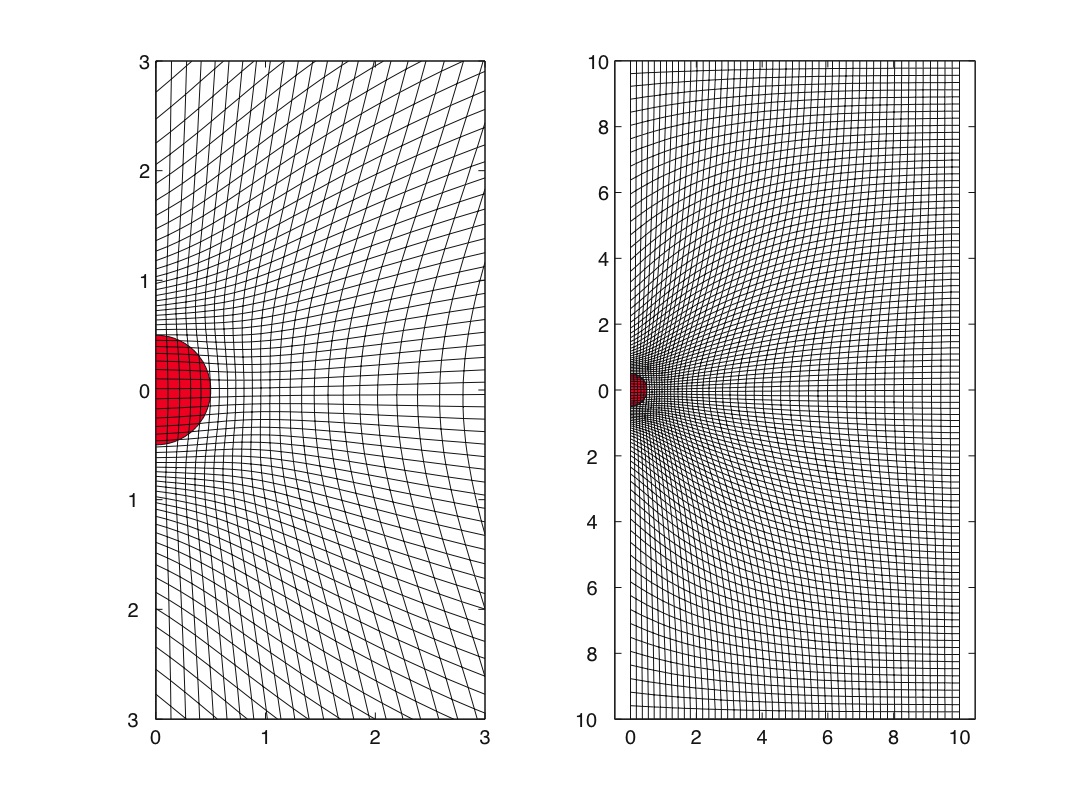} \caption{Initial setup of
 a dust charging simulation. The dust particle is represented by material
 computational particles with appropriate dielectric properties for the
 immersed boundary method. An adaptive grid is used to resolve the small
 sub-Debye scale dust particle.}
 \label{dustgrid}
\end{figure}

Figure~\ref{dustgrid} shows the configuration of the grid and of the
dust particle for the problem considered here. Note that a
nonuniform (but constant in time) grid is used to describe better
the sheath around the dust particle. The distance of the dust
particle from the boundaries is 10~$\lambda_{De}$. The plasma
species are initially loaded according to a drifting maxwellian
distribution with a downward vertical net flow velocity
corresponding to a Mach number $M=10$. To reach an equilibrium,
particles that flow out of the lower boundary are replaced by
particles injected at the top boundary~\cite{popdust99}.

\begin{figure}[htb]
\centering
 \includegraphics[width=80mm,angle=0]{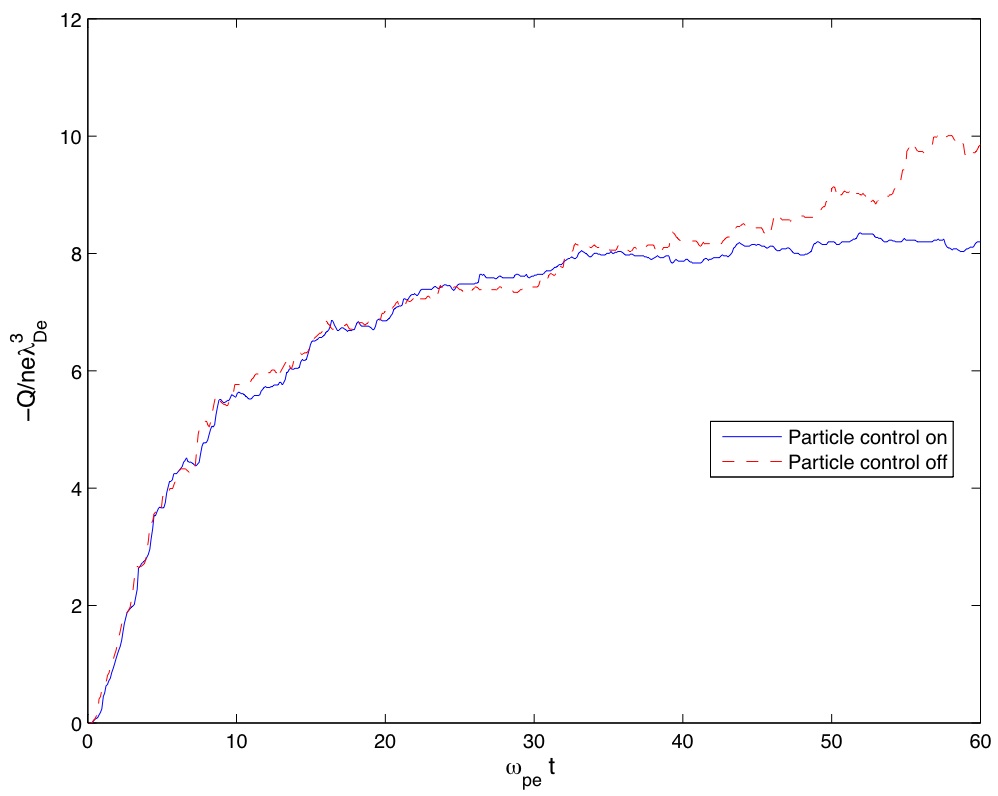} \caption{Evolution of the
 charge collected by the dust particle. Two runs are shown, both have uniform grids
 but one has also particle control (solid line) and the other has no particle control (dashed). }
 \label{dust}
\end{figure}
Figure~\ref{dust} shows the history of the net charge accumulated on
the dust particle. In this case, particle rezoning was used to
ensure the accuracy of the calculation. The particles are loaded,
initially, with a constant number of particles per cell, leading to
a higher concentration around the dust particle where the cells are
smaller. However, the plasma flow tends to empty the region around
the dust reducing the accuracy. Splitting the particles moving
toward the dust and coalescing the particles moving away from it is
desirable to keep the number of particles per cell and the accuracy
constant.

If the calculation is repeated without particle rezoning, the
accuracy worsens in time as the region around the dust becomes
less populated. Two effects lead to  decrease accuracy around the
dust particle: the particles originally present are in part
captured by the dust and in part just simply flow away according
to their average downward velocity of Mach $M=10$. The new
particles that replace them are flowing from regions of larger
cells and are less numerous leading to a decrease of accuracy.

As a result of the decrease in accuracy, the dust particle does not
reach a steady state in the run without particle rezoning.

\bibliography{lapenta}

\end{document}